\documentclass[aps,reprint, floatfix,longbibliography,superscriptaddress]{revtex4-1}
	\usepackage{graphicx}
	\usepackage{amsmath}
	\usepackage{amsfonts}
	\usepackage{amssymb}
	
\usepackage{braket}
	\usepackage{color}
    \usepackage[colorlinks=true, citecolor=blue, urlcolor=blue, linkcolor=black]{hyperref}
    
\makeatletter
\def\@bibdataout@aps{%
\immediate\write\@bibdataout{%
@CONTROL{%
apsrev41Control%
\longbibliography@sw{%
    ,author="08",editor="1",pages="1",title="0",year="1"%
    }{%
    ,author="08",editor="1",pages="1",title="",year="1"%
    }%
  }%
}%
\if@filesw \immediate \write \@auxout {\string \citation {apsrev41Control}}\fi 
}
\makeatother

\begin{document}

\title{A Cooper-Pair Box Coupled to Two Resonators: An Architecture for a Quantum Refrigerator}

\author{Andrew Guthrie}\email{andrew.guthrie@aalto.fi}
\author{Christoforus Dimas Satrya}\email{christoforus.satrya@aalto.fi}
\author{Yu-Cheng~Chang}
\address{Pico group, QTF Centre of Excellence, Department of Applied Physics, Aalto University School of Science, P.O. Box 13500,00076 Aalto, Finland}
\author{Paul Menczel}
\address{Theoretical Physics Laboratory, RIKEN Cluster for Pioneering Research, Wako-shi, Saitama 351-0198, Japan}
\author{Franco Nori}
\address{Theoretical Physics Laboratory, RIKEN Cluster for Pioneering Research, Wako-shi, Saitama 351-0198, Japan}
\address{Department of Physics, The University of Michigan, Ann Arbor, Michigan 48109, USA}
\author{Jukka P. Pekola}
\address{Pico group, QTF Centre of Excellence, Department of Applied Physics, Aalto University School of Science, P.O. Box 13500,00076 Aalto, Finland}
\address{Moscow Institute of Physics and Technology, 141700 Dolgoprudny, Russia}

\date{\today}

\begin{abstract}

Superconducting circuits present a promising platform with which to realize a quantum refrigerator. Motivated by this, we fabricate and perform spectroscopy of a gated Cooper-pair box, capacitively coupled to two superconducting coplanar waveguide resonators with different frequencies. We experimentally demonstrate the strong coupling of a charge qubit to two superconducting resonators, with the ability to perform voltage driving of the qubit at $\mathrm{GHz}$ frequencies. We go on to discuss how the measured device could be modified to operate as a cyclic quantum refrigerator by terminating the resonators with normal-metal resistors acting as heat baths.
\end{abstract} 
\maketitle

\section{Introduction}
The study of quantum heat engines and refrigerators plays a key role in the investigation of the fundamental relationship between quantum mechanics and thermodynamics \cite{annurev-physchem-040513-103724,Alicki_1979,PhysRevE.76.031105,Binder2019}. However, experimental realizations of cyclic quantum thermal engines have remained elusive. Such systems, in their most basic form, consist of a working substance with quantized energy levels which can be selectively coupled to a series of thermal reservoirs, and are capable of transporting heat \cite{e15062100}. By modulation of the working substance energy-level separation, for example, the system can be tuned to interact with each thermal reservoir sequentially. Moreover, by periodic modulation of the system energy levels, a quantum heat engine, or quantum refrigerator can be actualized \cite{PhysRevB.94.184503,PhysRevB.76.174523}.

A multitude of platforms have been proposed and explored to realize quantum thermal machines. In a seminal paper \cite{Rossnagel325}, an ion held in a linear Pauli trap was used to extract work by alternate exposure between a white noise electric field (hot reservoir), and a laser cooling beam (cold reservoir). More recently, a solid state quantum dot was operated as a `particle exchange' heat engine, where the dot can control a thermally driven flow of charge carriers \cite{Josefsson2018}. In a further development, a $^{13}$C nuclear spin has been utilized to implement an Otto cycle using a nuclear magnetic resonance setup \cite{PhysRevLett.123.240601}. Additionally, an electron spin impurity has been shown to act as an analogue heat-engine, with the `thermal' reservoirs inferred by the relative chemical potential in the leads \cite{PhysRevLett.125.166802}. However, in all such systems the heat current cannot be probed directly, and must be inferred from an additional parameter.

Circuit quantum electrodynamics (c-QED) using superconducting qubits remains a highly promising platform for realizing such a thermal machine, owing largely to the exceptional control which experimentalists have over the collective quantum degrees of freedom \cite{You2005,Gu2017,Kjaergaard2020}. c-QED has enjoyed a striking period of advancement, with numerous studies demonstrating strong coupling of photons to various types of qubits \cite{Wallraff2004, PhysRevA.69.062320,Schuster2007}, with a broad range of applications \cite{Clarke2008, Montanaro2016, Wendin2017}. Furthermore, modern nanofabrication techniques allow the integration and characterization of superconducting qubits coupled to normal-metal dissipative elements \cite{doi:10.1063/1.5098310}, creating hybrid c-QED systems capable of probing thermal transport in quantum systems. Such systems differentiate themselves from previous attempts on quantum heat engines via their unambiguous implementation of thermal reservoirs, which naturally define the bath temperature, and possess a multitude of techniques for both primary and secondary thermometry \cite{RevModPhys.78.217, pekola2021colloquium}.

\begin{figure*}[ht]
	    \includegraphics{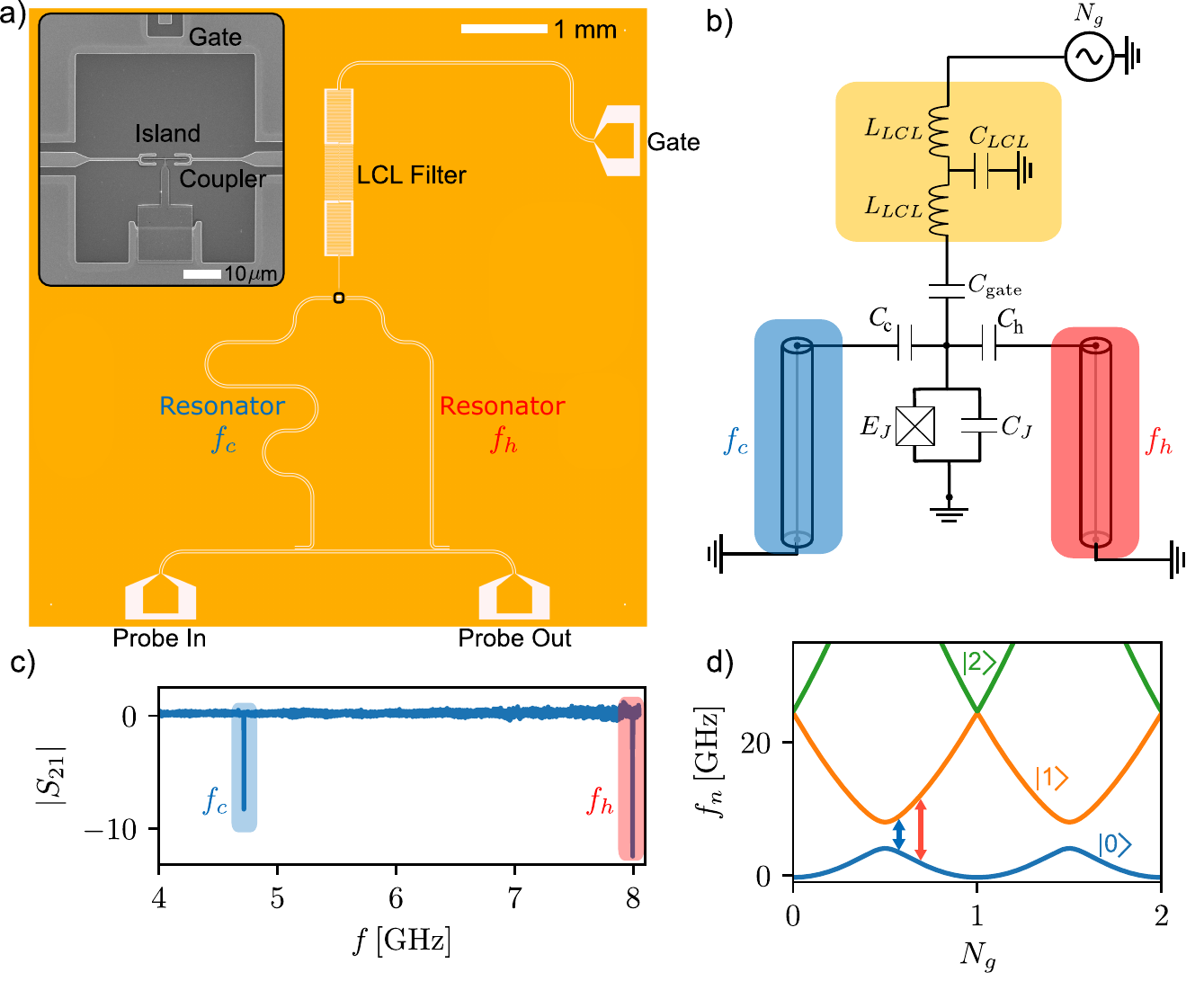}
	    \caption{(a) Design overview of the measured device. Two quarter-wavelength resonators of differing frequency are inductively coupled to a common feedline for readout. The voltage antinode of each resonator is capacitively coupled to a common superconducting island, connected to ground through a single Josephson junction (shown in the inset), and controllable by a nearby voltage gate. (b) Equivalent circuit for the measured sample with the feedline excluded. (c) Calibrated $|S_{21}(20~\mathrm{mK}) / S_{21}(4~\mathrm{K})|$ transmission through the feedline showing signals from two resonators and no spurious modes. (d) Calculated energy spectrum of the Cooper-pair box showing the first three energy levels for $E_c/h = 6.8~\mathrm{GHz}$ and $E_J/h = 3.5~\mathrm{GHz}$, clearly showing high anharmonicity around the degeneracy point. The blue and red arrows indicate the level spacing corresponding to the low and high frequency resonators respectively.
	    \label{fig:1}}
\end{figure*}

Superconducting quantum circuits involving dissipative elements have already platformed several pioneering experiments in quantum heat transport. A transmon qubit coupled to two superconducting resonators, terminated by normal metal resistors was used to measure DC heat transport modulated by magnetic flux threading a superconducting quantum interference (SQUID) loop. By using both identical and non-identical resonator frequencies, this led to the realization of a quantum heat valve \cite{Ronzani2018a} and a quantum heat rectifier \cite{Senior2020}. Despite the remarkable control exhibited by such systems, GHz-frequency cyclic driving of transmon qubits has proven experimentally difficult due to the large power dissipated by on-chip magnetic flux bias lines. Additionally, the performance of transmon qubits in thermal systems is limited by the relatively weak anharmonicity due to the large ratio of Josephson energy to charging energy, $E_J/E_C$. Weak anharmonicity removes the ability to properly isolate a qubit transition, meaning contributions from higher energy levels create undesired parasitic coupling \cite{Senior2020}. Moreover, the relatively long coherence time of a transmon qubit, a key asset in quantum information applications, can limit the performance of cyclic quantum engines due to the buildup of coherences in a process known as `quantum friction' \cite{PhysRevA.99.062103, Insinga2020}. The theoretical operation of qubits as thermal machines has been explored extensively, with promising proposals for implementing both refrigerators \cite{PhysRevApplied.16.044061, Abah_2016, PhysRevE.93.062134,PhysRevB.76.174523, PhysRevB.94.184503,PhysRevB.100.085405} and heat engines \cite{PhysRevB.93.041418, Campisi2016, PhysRevE.76.031105, PhysRevB.101.054513, PhysRevLett.93.140403, PhysRevE.87.012140}. 


To address the aforementioned problems to realize a cyclic quantum refrigerator a Cooper-pair box (charge qubit), which is isolated from higher energy states and easily modulated by an electric-field, could be used as a working substance. A charge qubit, in its simplest form, consists of a nanoscale superconducting island, grounded though a Josephson junction \cite{Nazarov2009}. Due to the small island dimension, the qubit frequency can now be tuned via the offset charge, $N_g$, on the island - controllable via the voltage of a nearby gate. Charge qubits have seen extensive study, both individually \cite{Nakamura1999, Pashkin2009, Bladh2005} and embedded in microwave cavities \cite{Sillanpaeae2006, Astafiev2007,Kim2008, Kim2011}. The strong coupling of a single photon to a charge qubit was demonstrated in a pioneering work of c-QED \cite{Wallraff2004}. Furthermore, modulated DC heat transport through a charge sensitive superconducting single electron transistor has been realized and explained with a simple theoretical model \cite{Maillet2020}. Charge qubits have experienced diminishing popularity in recent years since their charge sensitivity \cite{PhysRevA.76.042319} leads to high dephasing rates in quantum information applications. Nonetheless, their charge sensitive properties could be exploited to create an efficient working substance which can be operated with remarkably small input signals. We further note that a charge sensitive qubit connecting two cavity resonators could be a fundamental component in the field of quantum information processing. The setup could allow interactions between distant transmon qubits to be controlled using voltage gates rather than magnetic flux lines - significantly reducing the power required to realize two-qubit gates, compared to tunable flux qubits or SQUIDs \cite{Stassi2020}. 

In this work, we experimentally demonstrate a charge sensitive qubit capacitively coupled to two $\lambda/4$ resonators of differing frequency. We utilize a charge qubit consisting of an $8~\mathrm{nm}$ thick and $12~\mathrm{\mu m}$ long superconducting island connected to ground through a single-Josephson junction. Using a single junction, rather than the more common two junction SQUID approach, allows us to achieve higher $E_c$ whilst reducing the sensitivity of the qubit frequency to stray magnetic fields. Furthermore, we implement the ability for GHz driving of the qubit via an on-chip voltage gate in close proximity. Through a novel gate-line filtering scheme,  we can prevent microwave leakage from the resonators, and decay of the qubit, whilst maintaining the ability to drive at GHz frequencies. Our results show that, despite the sub-$\mu$m island dimensions, we can realize strong coupling of the qubit with two high-Q resonators without creating any spurious hybridized modes of the system. We go on to discuss how the measured device could be modified to operate as a quantum refrigerator, and perform numerical simulations using a Markovian master equation \cite{Lindblad1976,HeinzPeterBreuer2007}.

\section{Cooper-Pair Box coupled to Two Resonators}
We consider a charge qubit operating deep in the charge sensitive regime ($E_c \sim 2 E_J$) which couples with two resonators with different frequency. Here, the energy states correspond closely to the charge states, $|N\rangle$, on the island, and the Hamiltonian of the bare qubit is given by \cite{Nazarov2009}
\begin{align}\label{eq:H:qubit}
     \begin{split}
    {H}_Q = \sum\limits_{N} \Big[ 4E_c(N - N_g)^2 |N\rangle \langle N | 
    \\ - \frac{E_J}{2} (|N\rangle \langle N+1 | + |N+1\rangle \langle N |) \Big],
         \end{split}
\end{align}
where $N_g = C_\mathrm{gate} V_\mathrm{gate} /2e$ is the dimensionless offset charge controlled by the nearby gate voltage $V_\mathrm{gate}$, with capacitance $C_\mathrm{gate}$. At sufficiently low temperature and with restricted $N_g$ around the degeneracy point ($N_g = 0.5$), we can consider only two charge states: $|0\rangle$ and $|1\rangle$, and approximate the charge qubit Hamiltonian as a two-level system, described by
\begin{equation}\label{eq:H:qubit2}
    {H}_Q = -2E_c(1 - 2N_g) \sigma^z  -  \frac{E_J}{2} \sigma^x,
\end{equation}
where $\sigma^z, \sigma^x$ are the corresponding $2 \times 2$ Pauli matrices in the charge basis. The energy transition of the qubit will be
\begin{equation}\label{eq:qubitfrequency}
    \hbar \omega_Q=\sqrt{16E_{c}^2(1-2N_g)^2+E_{J}^2}. 
\end{equation}

The qubit interacts capacitively with both voltage antinodes of two quarter wavelength resonators. The Hamiltonian of each resonator is $H_{R,i} = h f_i a_i^\dag a_i$ for $i=\{c,h\}$, with the qubit-resonator interaction terms given by \cite{PhysRevA.69.062320}
\begin{equation}\label{eq:H:interaction}
    {H}_{I,i} =  g_{i,0}(a_i^\dag + a_i) [1 - 2N_g - \cos(\theta)\sigma^z + \sin(\theta) \sigma^x],
\end{equation}
where $\theta = \arctan(E_J/4E_C(1-2 N_g))$ is the qubit mixing angle. The atom cavity coupling at the degeneracy point is determined by the zero-point energy fluctuations of the cavity electric potential, and given by \cite{PhysRevA.76.042319} 
\begin{equation}\label{eq:g0}
g_{i,0} = e \frac{C_i}{C_\Sigma}\sqrt{\frac{hf_i}{l_i \tilde{c}}},
\end{equation}
where $C_i$, $C_\Sigma$ are the coupler-island and island total capacitances respectively, $l_i$ is the resonator length, and $\tilde{c}$ is the capacitance per unit length. By adding a term accounting for the resonator-resonator coupling, $\propto \tilde{g}$, the full system Hamiltonian is therefore 
\begin{equation}\label{eq:Hamiltonian}
    H =  H_Q + \sum\limits_{i = c,h}(H_{R,i} + H_{I,i}) + \tilde{g}(a_c^\dag a_h + a_h^\dag a_c).
\end{equation}
Operating close to the degeneracy point the full Hamiltonian can be reduced to a simplified two-cavity Jaynes-Cummings-type Hamiltonian. The effective coupling strength $g_i$, i.e., the coupling strength when operating away from the degeneracy point in resonance with the respective resonator, is reduced by a factor $\sin(\theta)$ in this framework. By implementing a time dependent $N_g(t)$ field, the qubit transition frequency $\omega_Q(t)$ can be cyclically modulated to interact back and forth between the two resonators $f_c$ and $f_h$.



    \begin{figure}[t]
    	    \includegraphics[width = \linewidth]{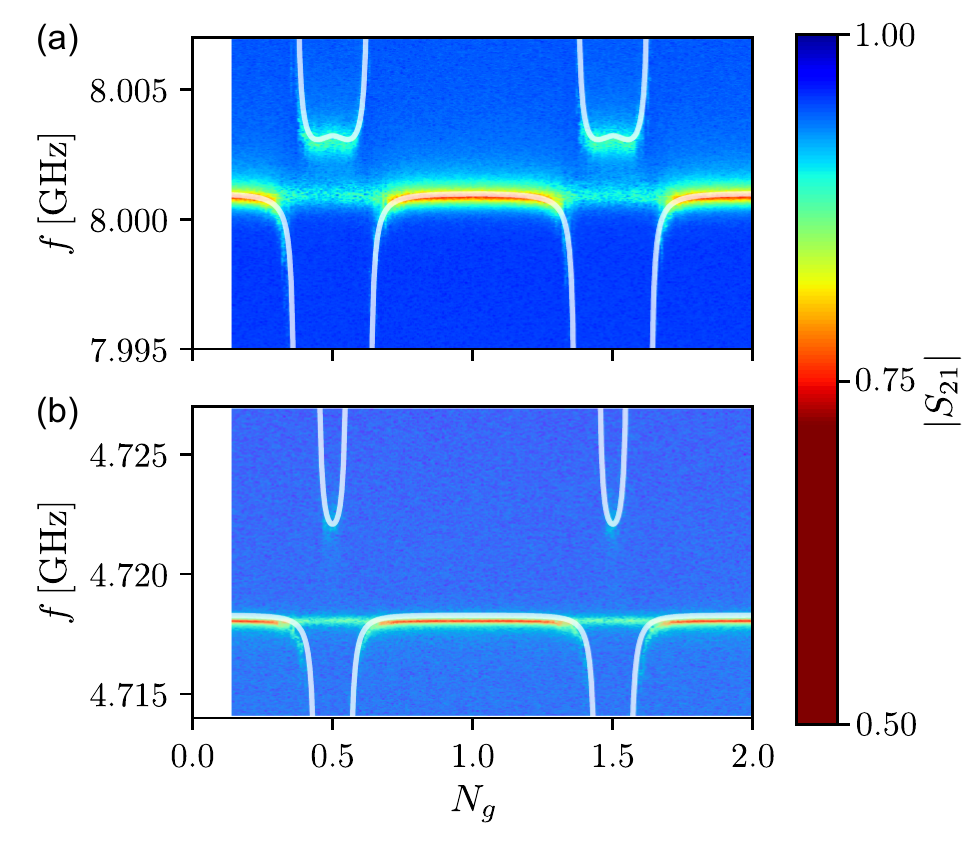}
    	    \caption{Spectroscopy of the resonator-qubit-resonator system, showing interaction of the charge qubit with both resonators. Avoided crossings are shown with (a) high-frequency resonator ($f_h$), and (b) low-frequency resonator ($f_c$). Results from both resonators were collected simultaneously to remove charge bias offsets. The solid white lines are simulated eigenenergies of the system, calculated using the SCQubits \cite{groszkowski2021scqubits} package to diagonalize the full Hamiltonian in Eq.~\ref{eq:Hamiltonian}. The model is calculated for $E_C/h = 6.8~\mathrm{GHz}$, $E_J/h = 3.5~\mathrm{GHz}$, $g_{c,0}/2\pi = 140~\mathrm{MHz}$,  $g_{h,0}/2\pi = 250~\mathrm{MHz}$, and $\tilde{g} = 0$.}
    	    \label{fig:2}
        \end{figure}

\section{Experimental Setup and Device}
The measured device consists of a charge qubit, formed by a superconducting island connected to ground through a tunnel junction, capacitively coupled to two quarter wavelength, coplanar waveguide (CPW) resonators. A render of the device layout, along with the equivalent circuit, is shown in Fig.~\ref{fig:1}(a) and Fig.~\ref{fig:1}(b). Figure~\ref{fig:1}(b) is depicted without the feedline. We fabricate the device on a $670~\mathrm{\mu m}$ thick and high-resistive silicon wafer. Ground-planes, feedlines, couplers and resonators are formed by reactive ion etching of a $200~\mathrm{nm}$ niobium (Nb) film. The qubit is then formed using a Dolan bridge technique to deposit two layers of aluminum (Al), with an in-situ oxidation to form the tunnel junction. The Nb surface is milled in-situ to ensure a clean contact between the Al and Nb film. Further detail on the fabrication procedure can be found in App.\,\ref{app:fabrication}. 

The inset of Fig.~\ref{fig:1}(a) presents a scanning electron micrograph of the qubit, showing the $12~\mathrm{\mu m}$ superconducting island, two Nb couplers and gate line. The Josephson energy, $E_J = 3.5~\mathrm{GHz}$ is controlled via the parameters of the oxidation, and can be estimated by measuring the normal-state resistance ($R_N = 42~\mathrm{k \Omega}$) of replica junctions fabricated alongside the main structures. By using fork-shaped coupling structures on either side of the small superconducting island we maximize the coupling strength by increasing the ratio $C_i/C_\Sigma$, as shown in Eq.~\eqref{eq:g0}. Using COMSOL simulations we estimate the capacitances to be $C_{c} = C_{h} = 460~\mathrm{aF}$, $C_\mathrm{gate} = 5~\mathrm{aF}$, and the measured $C_{\Sigma} = 2.7~\mathrm{fF}$, allowing a remarkable $C_{i}/C_\Sigma = 17\%$ to be achieved each, competitively large even when compared with single-resonator systems \cite{Astafiev2007,PhysRevA.76.042319}.

The two resonators are terminated to ground close to a common feedline, creating inductive coupling to allow excitation and readout. By coupling both resonators to a single feedline, we perform single-tone spectroscopy of the qubit in a wide frequency range, confirming the interaction of the qubit with each resonator. The resonators are read out through a notch-type measurement, by measuring the scattering parameter, $S_{21}$ \cite{Probst2015}. Both resonators are overcoupled to the feedline (i.e. the coupling quality factor $Q_c$ is less than the internal quality factor $Q_i$), to allow straightforward measurements in the single-photon regime. All measurements are performed in a cryogen-free dilution refrigerator, with a base temperature $20~\mathrm{mK}$.  A detailed diagram and further description of the measurement setup can be found in App.~\ref{app:setup}. Figure~\ref{fig:1}(c) shows the calibrated $S_{21}$ data as measured through the common feedline. Calibration is performed by measuring the same sample close to the critical temperature of the Nb film, and then correcting via plotting $|S_{21}(20~\mathrm{mK}) / S_{21}(4~\mathrm{K})|$. In this way, impedance mismatches due to the circuit are removed and only the temperature dependent resonator structures remain. We can identify two peaks at $f_c = 4.718~\mathrm{GHz}$ and $f_h = 8.001~\mathrm{GHz}$, corresponding to the two $\lambda/4$ resonators. We note the absence of any parasitic or hybridized modes, suggesting the resonator-resonator cross-talk is small.

\section{Spectroscopy of the Resonator-Qubit-Resonator System} \label{sec:spec}
To confirm the interaction of the qubit with each resonator, we perform simultaneous one-tone spectroscopy of the resonator-qubit-resonator system. We use a vector network analyzer (VNA) to probe a frequency, $f_\mathrm{probe}$, in the vicinity of the two resonator frequencies, whilst varying the DC gate voltage. The number of photons in the cavity is estimated to be less than $1$ in both cases. Figure\ \ref{fig:2} shows the one-tone spectra in the proximity of the two frequencies, as a function of the dimensionless offset charge, $N_g$. The Rabi splitting associated with the interaction of a two-level system with a cavity is shown clearly for the high-frequency resonator (Fig.~\ref{fig:2}(a)) and the low-frequency resonator (Fig.~\ref{fig:2}(b)). Two periods are presented, symmetrically around $N_g = 1$. The solid white lines are fits using the numerical solutions to  Eq.~\eqref{eq:Hamiltonian}, with $E_C/h = 6.8~\mathrm{GHz}$, $E_J/h = 3.5~\mathrm{GHz}$, with excellent agreement with the theoretical model.

In our Cooper-pair box, the spatial profile of the superconducting gap energy is controlled by a thickness difference between the Al-island and Al-lead to suppress the quasiparticle-tunneling rate across the junction \cite{Yamamoto2006}. As a result, we observe one-Cooper-pair periodicity of the Rabi splitting in the one-tone spectroscopy. Nevertheless, the poisoning is still expected to be present close to the degeneracy point $N_g = 0.5$ \cite{Aumentado2004}. This poisoning can be observed by the presence of some leftover signal at the cavity frequency in both cases, providing some remaining off-resonance qubit signal \cite{PhysRevLett.108.230509,PhysRevApplied.12.014052}. We observe this averaging effect because the average parity-switching rate is much shorter than the measurement time. By comparison of the relative amplitudes of the bare resonance signal with that of the remaining signal at the degenerecy we can estimate the parity preference for the odd and even states. We find an even state preference of $66 \%$ and $61 \%$ in the case of the high-frequency and low-frequency resonators respectively. This could be further mitigated through improved infrared shielding and further quasiparticle engineering \cite{catelani2021using}. In fact, quasiparticle tunneling is the dominant source of longitudinal relaxation, further discussed in Sec.~\ref{app:twotone}.

The relative coupling strengths are measured to be $g_{c,0}/2\pi = 140~\mathrm{MHz}$,  $g_{h,0}/2\pi = 250~\mathrm{MHz}$ and $\tilde{g} \sim 0$. The negligible crosstalk is apparent in Fig.~\ref{fig:1}(c) from the lack of hybridized modes in the $S_{21}$ spectrum, and expected due to the vanishing spectral overlap of the two resonators' Lorentzian functions. We note the relatively weak signals for the dressed states as they move far from the cavity frequency, indicative of the higher dissipation associated with charge sensitive devices when compared with transmon type qubits. Although the raw couplings, $g_{c,0}$, $g_{h,0}$, are large, due to the charge sensitivity the effective couplings are reduced by a factor of $\sin{\theta}$, and decrease as we move away from the degeneracy point. We extract the effective coupling strengths by measuring the dispersive shift, $\chi_i$, of the resonance at the degeneracy point and calculate by $g_i = \sqrt{\chi_i \Delta_i}$, where $\Delta_i = (f_i - E_J/h)$ is the detuning at degeneracy. The effective coupling strengths are $g_{h}/2\pi = 125~\mathrm{MHz}$ and $g_{c}/2\pi  = 76~\mathrm{MHz}$, corresponding to $1.6\%$ and $2\%$ of the resonance frequency respectively. Furthermore, the extracted $E_J$ and $E_C$ can be confirmed experimentally using a two-tone spectroscopy technique, to probe the exact qubit transition in the vicinity of the degeneracy point, discussed in detail in Sec.~\ref{app:twotone}.


Importantly, due to the large $E_c$ and $C_\mathrm{gate}$ in the measured device, we can achieve qubit control in a large frequency range using remarkably small signals. Based upon the measured system parameters, the qubit could be driven sinusoidally between the two resonators using a signal amplitude $V_\mathrm{rms} = 1.2~\mathrm{mV}$, corresponding to just $10~\mathrm{nW}$ power at $50~\mathrm{\Omega}$. This presents a four orders-of-magnitude improvement over a comparable driving scheme using an on-chip magnetic flux bias line. The presented data is, to the authors knowledge, the first use of a \emph{charge sensitive} qubit as a coupling element between two superconducting resonators.

\section{Decoherence of the Charge Qubit} \label{app:twotone}

    \begin{figure}[t]
    	    \includegraphics[width=\linewidth]{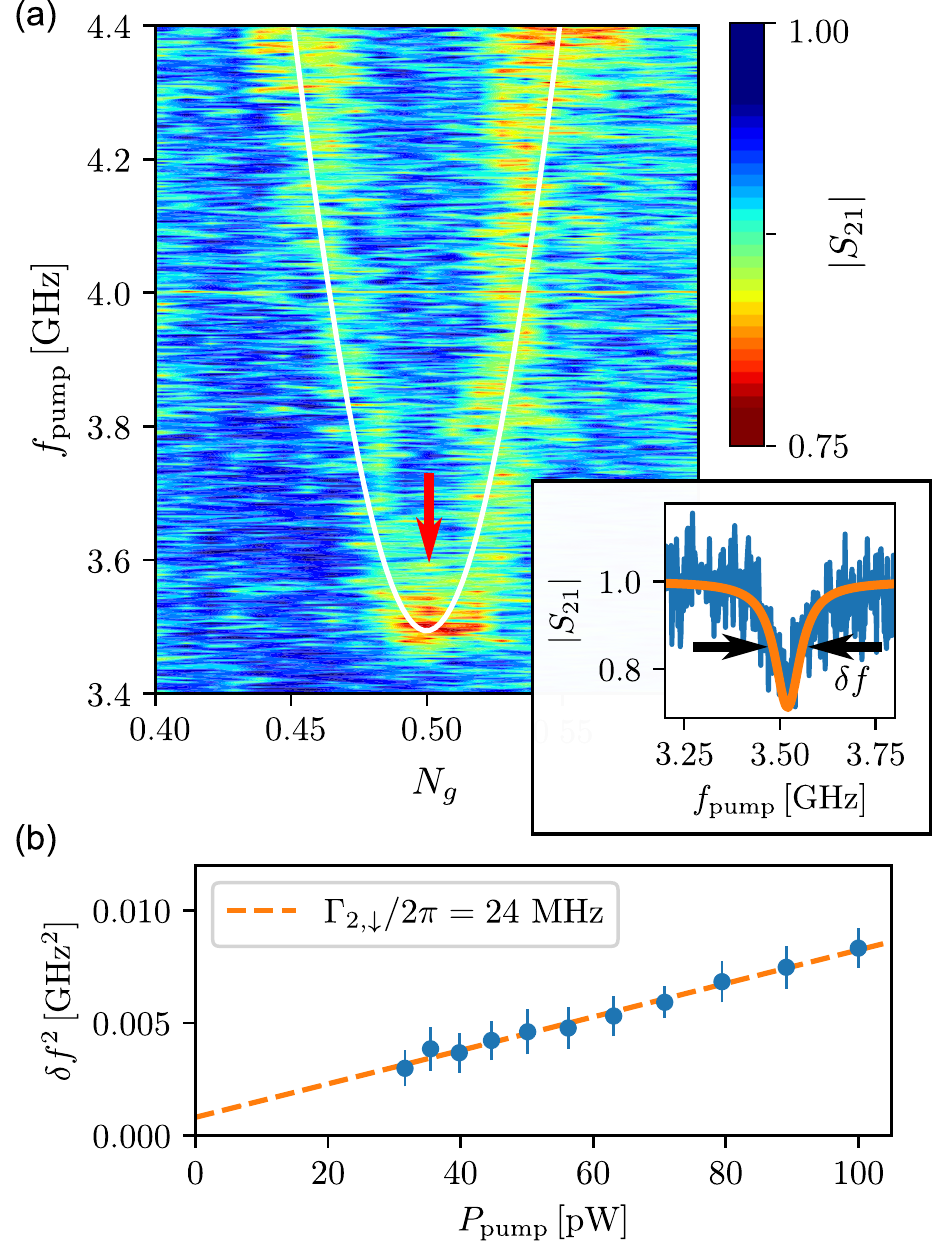}
    	    \caption{(a) Two-tone spectroscopy of the charge qubit in the vicinity of the degeneracy point ($N_g = 0.5$), with fixed probe tone $f_\mathrm{probe} = 8.001\,\mathrm{GHz}$, and sweeping pump tone $f_\mathrm{pump}$ from 3.4 to 4.4~GHz. The solid white line indicates the transition energy of the two-level system qubit solved from the Hamiltonian Eq.~\eqref{eq:Hamiltonian} using SCQubits, for $E_c/h = 6.8\,\mathrm{GHz}$ and $E_J/h = 3.5\,\mathrm{GHz}$. The inset shows a frequency slice of 2D data, as indicated by the red arrow, and is fitted using a Lorentzian function. (b) Squared spectral linewidth of the Lorentzian as a function of pump power, $P_\mathrm{pump}$, averaged by repeated sampling of the curves. The dashed orange line presents a fit using Eq.~\eqref{eq.gamma2}. By extrapolation ($P_\mathrm{pump} \to 0$, $n_p \to 0$), we estimate  the dephasing time to be $\Gamma_{2, \downarrow} / 2 \pi = 24~\mathrm{MHz}$.}
    	    \label{fig:5}
        \end{figure}

In order to quantify the qubit decoherence and to map the two lowest energy levels of the Cooper-pair box, two-tone spectroscopy is performed. The transmission, $S_{21}$, of a weak microwave signal (probe signal) is continuously measured by a VNA located at room temperature. The system is probed at the bare-resonance frequency of the high-frequency resonator ($f_\mathrm{probe} = 8.001\,\mathrm{GHz}$), which is sensitive to the qubit-population. The second signal (pump signal) is generated by a built-in second generator of the same VNA, and combined using a signal splitter. When the pump frequency is in resonance with the qubit frequency, the qubit is excited and the measured probe $S_{21}$ drops. By repeating this procedure at different $N_g$, the qubit energy spectrum can be traced. Figure~\ref{fig:5}(a) is the spectrum obtained and fitted well by the $f_{01}$ transition calculated from the Hamiltonian \eqref{eq:Hamiltonian} using SCQubits, obtaining $E_c/h$=6.8~GHz and $E_J/h =3.5~\mathrm{GHz}$, in agreement with the results obtained in Sec.~\ref{sec:spec}. The inset in Fig.~\ref{fig:5}(a) shows a slice at $N_g = 0.5$ (blue line) and is fitted by a Lorentzian function (orange line) with a linewidth $\delta f$. 

The population of the excited state under continuous pumping, $p_1$,  can be found from the steady state solution to Bloch equations, and is given by \cite{PhysRevLett.94.123602}
\begin{equation}\label{eq:bloch11}
p_1 = \frac{1}{2 }\frac{4 n_p g_h^2/ (\Gamma_{1, \downarrow}\Gamma_{2, \downarrow})}{1 + ((f_Q - f_\mathrm{pump}) / \Gamma_{2, \downarrow})^2 + {4 n_{p}g_{h}^{2}}/{\Gamma_{1, \downarrow}\Gamma_{2, \downarrow}}},
\end{equation}
where $g_h$ is the coupling strength to the readout resonator, in this case the high-frequency resonator. By fitting with a Lorentzian, we find that the spectral linewidth relates to the longitudinal-relaxation rate, $\Gamma_{1, \downarrow}$, and phase-decoherence rate, $\Gamma_{2, \downarrow}$, of the qubit by
\begin{equation}\label{eq.gamma2}
2\pi\delta\!f=\Gamma_{2, \downarrow}\sqrt{1+\frac{4 n_{p}g_{h}^{2}}{\Gamma_{1, \downarrow}\Gamma_{2, \downarrow}}},
\end{equation}
where $n_p$ is the pump photon number. 

Figure~\ref{fig:5}(b) presents the dependence of the spectral linewidth squared for varying pump power, $P_\mathrm{pump}$, showing the expected power dependence, as $n_p$ $\propto$ $P_\mathrm{pump}$. The dashed-orange line shows a fit to the data using Eq.~\eqref{eq.gamma2}, allowing us to extract the spectral linewidth as $P_\mathrm{pump} \to 0$ by extrapolation. Here, due to the low power of the pump signal, $n_p \to 0$, the linewidth $\delta f$ is dominated by qubit dissipation \cite{PhysRevLett.94.123602}, thus $2\pi\delta f\sim\Gamma_{2, \downarrow}$, which is the qubit decoherence rate. The dissipation measured by this approach can be decomposed to the two relaxation processes, longitudinal relaxation, $\Gamma_{1, \downarrow}$, and pure dephasing, $\Gamma_\varphi$, related through the expression
\begin{equation}
\Gamma_{2, \downarrow} = \frac{\Gamma_{1, \downarrow}}{2}+ \Gamma_\varphi.
\end{equation}
Although our measurement technique does not allow us to extract the relative size of each contribution, previous experiments suggest that longitudinal relaxation is dominant close to the degeneracy point, where quasiparticle tunneling is the major contributor \cite{PhysRevLett.93.267007}.


\section{Voltage Driving and Suppression of Microwave Leakage}
   
     \begin{figure}[b] \includegraphics[width=\linewidth]{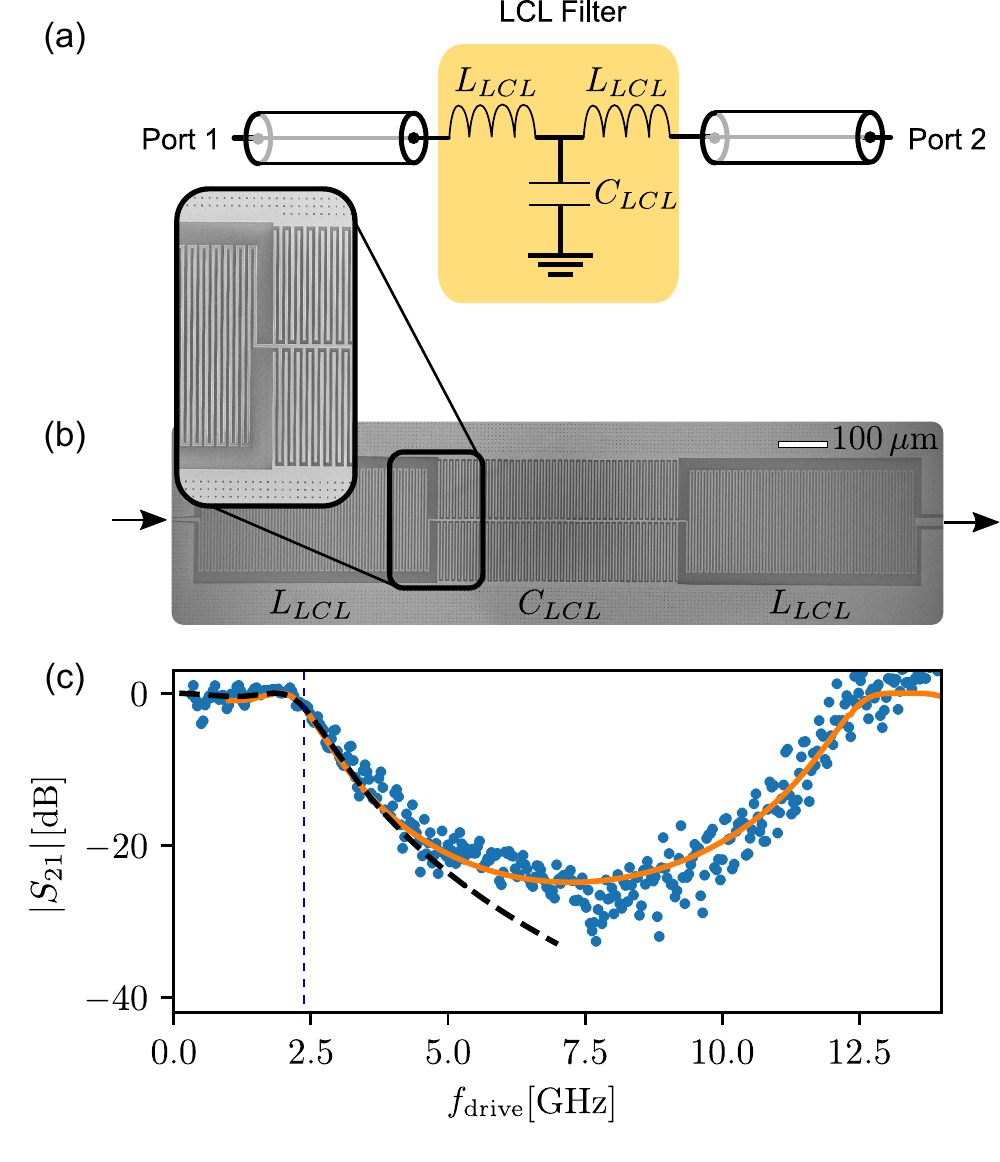}
    	    \caption{(a) Schematic lumped circuit of $LCL$-filter consisting of two series inductors grounded in the middle by a capacitor. (b) Scanning electron microscope (SEM) image of the fabricated $LCL$-filter comprising of two Nb-film meandering-line inductors and an interdigitated capacitor. (c) Transmission spectra, $S_{21}$, between port~1 and port~2 for driving freqencies $f_{\textrm{drive}}$ 0.1-14~GHz. Blue dots are measured data at $20~\mathrm{mK}$, the solid orange line and black dashed line are the SONNET simulation and analytical model respectively. The $LCL$ cutoff frequency, $f_\mathrm{cutoff}= 2.3~\mathrm{GHz}$, is shown by the blue dashed line.
    	    \label{fig:3}}
        \end{figure}   
     In order to drive the qubit energy cyclically, sinusoidal or arbitrary voltage driving should be applied through the gate line to modulate the qubit energy level between the low-frequency ($f_{c}$) and high-frequency ($f_{h}$) resonator, without introducing noise or microwave leakage through the driving line. Quarter-wave ($\lambda/4$) resonators, which have a voltage maximum at the open side, interact capacitively with the gate line, introducing some microwave leakage to the line. Filtering the operating range of the qubit (4-8~GHz) whilst allowing an AC-signal up to few GHz is pivotal. We utilize a superconducting $LCL$-circuit acting as a lowpass filter \cite{OriginalTFilter}, enabling us to prevent microwave leakage with around $20~\mathrm{dB}$ attenuation and to drive the qubit up to a cutoff frequency of few GHz.
     Figure~\ref{fig:3}(a) shows the schematic circuit of an $LCL$-filter consisting of two series inductors shunted at the center by a capacitor. In the fabricated device, as shown in Fig.~\ref{fig:3}(b), the filter is realized by a meandering-line inductor ($L_{LCL}$) and an interdigitated capacitor ($C_{LCL}$) both with a central width and line-spacing of $4~\mathrm{\mu m}$.
     
     To understand the transmission properties, the filter is separately characterized at $20~\mathrm{mK}$, as shown in Fig.~\ref{fig:3}(c), by measuring $S_{21}$ of a sinusoidal signal with frequency $f_{\textrm{drive}}$ between port~1 and port~2. The blue dots in Fig.~\ref{fig:3}(c) show the calibrated $S_{21}$ signal, and exhibit close to $100\%$ transmission up to $\sim2.3$~GHz and $>20$~dB attenuation within the range of 4-10~GHz. Above 10~GHz, the attenuation starts to decrease and reaches 0~dB at $\sim13~\mathrm{GHz}$. A lumped circuit model of transmission derived from the ABCD matrix (dashed black line), discussed in detail in App.~\ref{app:LCLFilter}, predicts the filter's behavior up to $\sim$~4~GHz and deviates above it due to the parasitic capacitance of the meandering inductors, which is not taken into account in the model. From the fitting, the cut off frequency $f_\mathrm{cutoff}= \sqrt{2/L_{LCL} C_{LCL}}$ is obtained to be 2.3~GHz with $L_{LCL}=5.9$~nH and $C_{LCL}=1.7$~pF. A finite element simulation using SONNET (orange line) captures fully the filter's transmission with excellent agreement. The characterized filter is capable of blocking microwave leakage within the frequency range of the two resonators, whilst still allowing cyclic driving of the qubit up to 2.3~GHz.

\section{Proposed Operation of the Quantum Otto Refrigerator}\label{sec:performance}

The measured device could be modified to operate as a quantum Otto refrigerator by terminating the two resonators by normal-metal resistors \cite{PhysRevB.94.184503}. The resistor-terminated resonators act as a thermal bath \cite{Ronzani2018a, Senior2020}. Furthermore, by connecting superconducting probes to the resistors through an insulating barrier the temperature of the two baths could be controlled and monitored through voltage and current bias respectively \cite{RevModPhys.78.217}. Due to the dissipation created by the normal-metal probes, the quality factor of the resonators would be very low (Q~$\sim 10$) \cite{doi:10.1063/1.5098310}, and qubit spectroscopic characterization could no longer be performed.

The Otto refrigerator cycle is the most practically achievable implementation of a quantum refrigerator. The Otto cycle consists of sequential interactions between a two-level system and a cold ($f_c$) and hot ($f_h$) reservoir. It has four branches: an adiabatic stroke of the qubit frequency from $f_c$ to $f_h$, thermalization with the hot bath at frequency $f_h$, an adiabatic stroke back from $f_h$ to $f_c$, and finally thermalization with the cold bath at frequency $f_c$. Cooling is achieved under the condition $f_h/f_c > T_h/T_c$, where $T_h$, $T_c$ are the temperatures of the normal-metal elements shunting the hot and cold resonators to ground, respectively.

      \begin{figure}[t] \includegraphics[width=\linewidth]{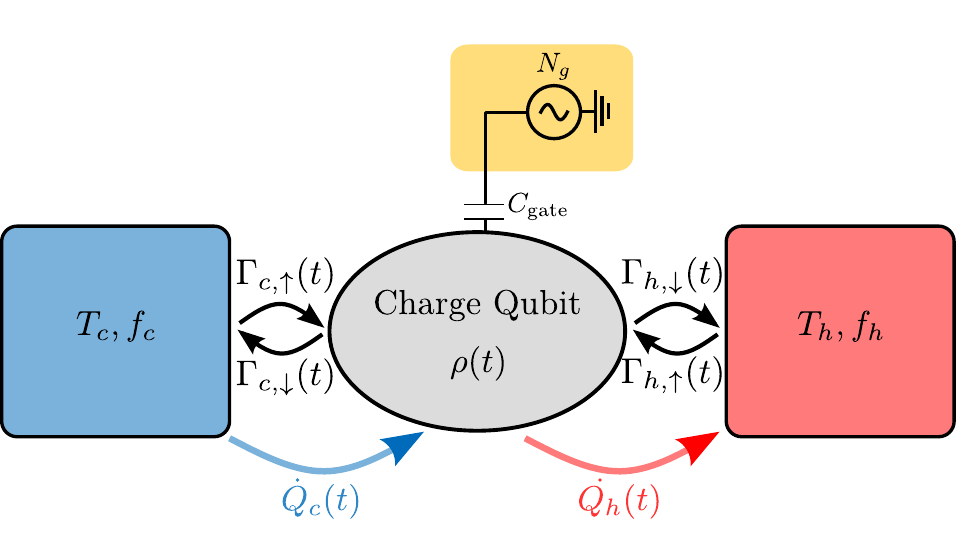}
    	    \caption{Equivalent heat flow diagram used for the numerical simulations of the quantum Otto refrigerator. We consider the case of two heat baths corresponding to the two resistor-terminated resonators. \label{fig:thermal}}
        \end{figure}

To implement the quantum Otto cycle in our system, we should consider Eq.~\eqref{eq:H:qubit2} however allowing time dependence in the offset charge $N_g(t)$. We consider driving the system with a truncated trapezoidal shape, 
     \begin{equation}
         q(t) = \frac{1}{2}\left[1 + \frac{\tanh(a \cos(2 \pi f_\mathrm{drive} t))}{\tanh(a)} \right],
     \end{equation}
     where $a$ is a constant, a form previously demonstrated to yield a large cooling power \cite{PhysRevB.94.184503}. The offset charge is applied by gate voltage, and driven in the form
         \begin{equation}
         N_g(t) = N_{g,c} +  (N_{g,h} - N_{g,c})q(t),
     \end{equation} 
     in which $N_{g,c}$ and $N_{g,h}$ are the offset charge at which the qubit interacts with the cold and hot reservoir respectively, obtained from rearranging Eq.~\eqref{eq:qubitfrequency}.

    For a qubit coupled to a resonator its emission rates are altered according to the Purcell effect, and can be calculated using Fermi's golden rule \cite{PhysRevA.76.042319}. Following \cite{PhysRevB.94.184503, PhysRevB.100.035407} we consider the dissipators to take the form of Johnson-Nyquist noise generated by a normal metal resistor, and spectrally filtered by the Lorentzian function of the resonators. The transition rates are therefore described by
     \begin{equation}
         \Gamma_{i, \downarrow}(t) = \frac{g_i}{4 \pi}\frac{1}{1 + Q_i^2 ({\omega_i}/{\omega_Q} - {\omega_Q}/{\omega_i})^2} \frac{\omega_Q}{1 - e^{{- \hbar \omega_Q}/{k_B T_i}}},
     \end{equation}
        where $\omega_i = 2 \pi f_i$ is the resonator frequency, $Q_i$ is the associated quality factor, $\omega_Q(t)$ is the instantaneous qubit frequency, and $T_i$ is the temperature of each normal-metal resistor terminating the resonator. The rates coming from the two heat baths obey the detailed balance condition as
      \begin{equation}
         \Gamma_{i, \uparrow}(t) = \Gamma_{i, \downarrow}(t)  \exp({{-\hbar \omega_Q}/{k_B T_i}}).
     \end{equation}  
In this way, the transition rates from each bath are maximized when the qubit is in resonance with the corresponding resonator frequency. Due to the finite quality factor of the resonators, this protocol can only approximate the Otto cycle since the qubit is never fully decoupled from either bath. Furthermore, the cooling power at the highest drive frequencies is limited by these transition rates, since the cycle becomes too short for the qubit to reach equilibrium with the resonators.




In limit where the bath-resonator coupling exceeds the resonator qubit coupling (local limit), and in the limit of slow driving,  the evolution of the qubit density matrix can be described by a Lindblad master equation as \cite{Lindblad1976}
     \begin{align} \label{lindblad}
     \begin{split}
        \dot{\rho(t)} = \frac{-i}{\hbar}\left[H_Q , \rho(t) \right] 
         + \Gamma_\uparrow(t)\left[\sigma_+ \rho(t) \sigma_- - \frac{1}{2}\{ \sigma_- \sigma_+ , \rho(t) \}
         \right] \\
         + \Gamma_\downarrow(t)\left[\sigma_- \rho(t) \sigma_+ - \frac{1}{2}\{ \sigma_+ \sigma_- , \rho(t) \} \right],
         \end{split}
         \end{align}
    where $\Gamma_{\uparrow/ \downarrow}(t) = \Gamma_{c, \uparrow/ \downarrow}(t) +\Gamma_{h, \uparrow/ \downarrow}(t)$ are the qubit transition rates, $\sigma_{\pm}$ are the instantaneous jump operators of the system and $\{a,b\}$ defines the anti-commutator operation.


 The equivalent heat flow diagram, including two heat baths and the corresponding rates, is shown in Fig.~\ref{fig:thermal}. We simplify Eq.~\eqref{lindblad} by transforming to the rotating frame using $\Bar{\rho} = V(t)^\dag \rho V(t)$ \cite{PhysRevB.99.224306}, where $V$ is the unitary matrix diagonalizing $H_Q(t)$. Then, parameterizing in terms of the Bloch equation elements, $x(t)$, $y(t)$, $z(t)$, the evolution results in the compact expressions
    \addtolength{\jot}{1em}
    \begin{widetext}
         \begin{align}
         \dot{x}(t) =  -\frac{\Gamma_\downarrow(t) + \Gamma_\uparrow(t)}{2} x(t) - \omega_Q(t)y(t) - \frac{8 E_c E_J \dot{N_g}(t)}{\hbar^2 \omega_Q^2} z(t),
         \\\dot{y}(t) =   \omega_Q(t)x(t) -\frac{\Gamma_\downarrow(t) + \Gamma_\uparrow(t)}{2} y(t),
         \\\dot{z}(t) =  \frac{8 E_c E_J \dot{N_g}(t)}{\hbar^2 \omega_Q(t)^2} x(t) - [\Gamma_\downarrow(t) + \Gamma_\uparrow(t)]z(t) - [\Gamma_\downarrow(t) - \Gamma_\uparrow(t)].
    \end{align}
    \end{widetext}
    
Based on this, we further note that the condition for adiabatic evolution in our system is \cite{PhysRevA.73.062307}
\begin{equation}
\frac{\braket{0|\dot{H}|1}}{h^2\omega_Q^2} \ll 1,
\end{equation}
where fulfilment means that the system remains in an instantaneous eigenstate throughout the evolution. 
Furthermore, we can write the exact heat currents from each bath in terms of the elements of the Bloch vector as 
         \begin{equation}
         \dot{\mathcal{Q}_i}(t) = \frac{-\hbar \omega_Q(t)}{2}\left( [\Gamma_{i,\downarrow(t)} + \Gamma_{i,\uparrow(t)}]z(t) + [\Gamma_{i,\downarrow(t)} - \Gamma_{i,\uparrow(t)}]\right),
     \end{equation}

     \begin{figure}[t] \includegraphics[width=\linewidth]{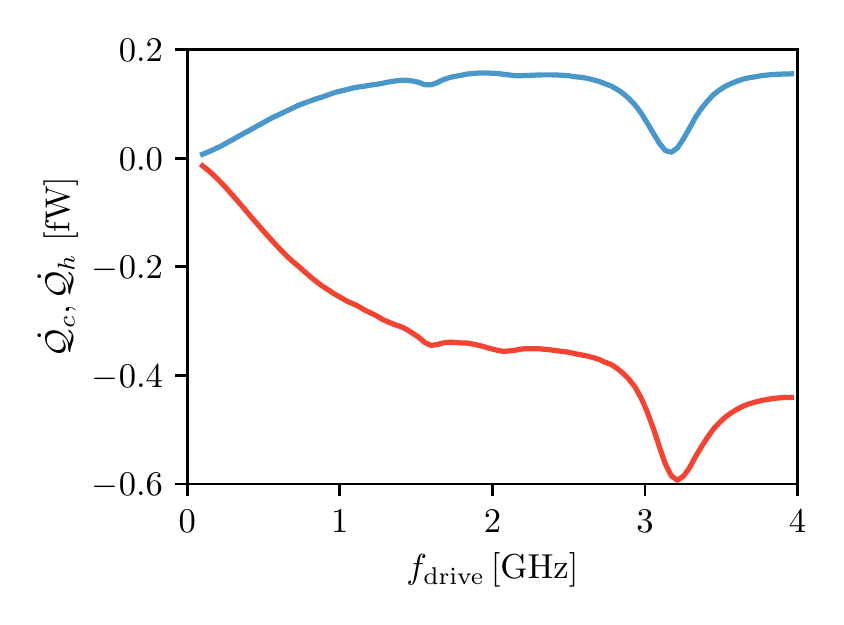}
    	    \caption{Numerical simulations of cooling power in a Cooper-pair box acting as a quantum Otto refrigerator for realistic values of the resonators quality factor $Q_c = Q_h = 2$ \cite{Ronzani2018a,Senior2020,doi:10.1063/1.5098310}, $T_c = T_h = 300~\mathrm{mK}$, and with $a=2$. The rising (blue) curve shows the heat extracted (positive) from the cold bath, and the falling (red) curve shows the heat extracted (negative) to the hot bath. At the highest drive frequencies we could achieve $\sim 150~\mathrm{aW}$ of cooling power, detectable using standard normal metal-insulator-superconductor thermometry techniques. \label{fig:4}}
        \end{figure}   

Figure~\ref{fig:4} shows the average power extracted from the hot and cold baths $\langle\dot{\mathcal{Q}}_i \rangle = \frac{1}{\tau}\int_0^\tau \dot{\mathcal{Q}}_i(t) dt$ as the red and blue solid lines for the measured system parameters, averaged over one cycle one the system has reached the steady state. Furthermore, it can be seen that the rate of entropy production is consistent according to the second law $\dot{S} - \sum_i \mathcal{\dot{Q}}_i / k_\mathrm{B} T_i \ge 0$.

Interestingly, the driving rate in our system could be high enough to observe quantum behavior in the refrigerator cooling power, whereby off-diagonal terms in the density matrix, $\rho$, could begin to affect the refrigerator performance. In the simulations, the quantum effects are clearly visible by sharp oscillations in the cooling power at high values of $f_\mathrm{drive}$, as has been seen in previous theoretical studies involving qubit Otto refrigerators \cite{PhysRevB.94.184503}. Dips are created when the frequency of the free qubit rotation about the Bloch sphere matches the driving frequency. In the future, to suppress this behavior, a counter-diabatic driving protocol could be implemented, which generally consists of an additional field to `guide' the qubit along an adiabatic trajectory \cite{PhysRevB.100.035407}. We note however, that for very fast driving the Lindblad operators may no longer be jump operators between the instantaneous eigenstates of the Hamiltonian. The modeling could be further improved by utilizing more advanced Floquet master equations, as was done in \cite{PhysRevE.87.012140}, for example.

In addition to the direct heat-flows discussed here, tunneling quasiparticles in the system will also affect the refrigerator performance. We see experimental evidences of this effect from the non-interacting portion of the signal seen in Fig.~\ref{fig:2}(a) and Fig.~\ref{fig:2}(b). Since each tunnelling quasiparticle shifts the qubit transition off-resonance, we can understand this effect as a reduction in the effective coupling strength to each resonator, i.e. $\bar{g}_c = 0.61 g_c$ and $\bar{g}_h = 0.66 g_h$. Incorporating such an effect into the numerics reduces the peak cooling power from $150~\mathrm{aW}$ to $60~\mathrm{aW}$, still detectable using standard normal metal-insulator-superconductor (NIS) thermometry techniques. However, this highlights the importance of further quasiparticle mitigation strategies.

\section{Conclusion}

In summary, a charge sensitive qubit has been coupled to two superconducting coplanar waveguides for the first time, with the ability to drive the qubit over a large frequency range using remarkably small excitations. Additionally, the measured effective coupling strength of the qubit to each resonator remains exceptionally high, competitive with the highest previously measured in charge sensitive devices. Furthermore, we demonstrate that despite the close proximity of the various coupling elements, our system can be simply described within the framework of a two-level qubit interacting with two resonators. Utilizing the measured device parameters, we propose and simulate the operation of our device acting as a quantum Otto refrigerator, and show cooling powers of the order $\sim 60~\mathrm{aW}$ which is detectable using normal metal-insulator-normal metal
(NIS) thermometry. Additionally, the measured system could be used to realize a highly effective heat rectifier, owing to the large anharmonicity of the charge qubit, allowing the isolation of a single qubit transition. Our work lays the technical foundation towards the realization of cyclic quantum heat engines within the c-QED framework, and opens the door towards a multitude of future studies in the field of quantum thermodynamics.

\appendix

\section{Fabrication Details} \label{app:fabrication}
The fabrication of the device is done in a multistage process on a $675~\mathrm{\mu m}$-thick and highly resistive silicon substrate. The fabrication consists of two main steps: patterning microwave structures on a Nb film, and Josephson-junction elements on an Al film. A $40~\mathrm{nm}$-thick~Al\textsubscript{2}O\textsubscript{3} layer is deposited onto a silicon substrate using atomic layer deposition, followed by a deposition of a $200~\mathrm{nm}$-thick Nb film using DC magnetron sputtering. Positive electron beam resist, AR-P6200.13, is spin-coated with a speed of 5500~rpm for 60~s, and is post-baked for 9 minutes at 150$^{\circ}$C, which is then patterned by electron beam lithography (EBL) and etched by reactive ion etching. A shadow mask defined by EBL on a 1~{$\mathrm{\mu m}$}-thick poly(methyl-metacrylate)/copolymer resist bilayer is used to fabricate the Al island and Josephson junction using a two-angle deposition technique at 0$^{\circ}$ and 32$^{\circ}$ sequentially. Before the deposition, the Nb surface is cleaned in-situ by Ar ion plasma milling for 45~s, followed by first 8~nm-thick Al island deposition. The island then is oxidized at pressure 2.5~mbar for 2.5~minutes to form a tunnel barrier before depositing the second 100~nm Al film. Finally, after liftoff in acetone and isopropyl alcohol, the substrate is cut by an automatic dicing-saw machine to the size $7\times7$~mm and wire-bonded to an RF-holder for the low-temperature characterization.

\section{Experimental Details}\label{app:setup}

    \begin{figure}[t]
    	    \includegraphics[width=\linewidth]{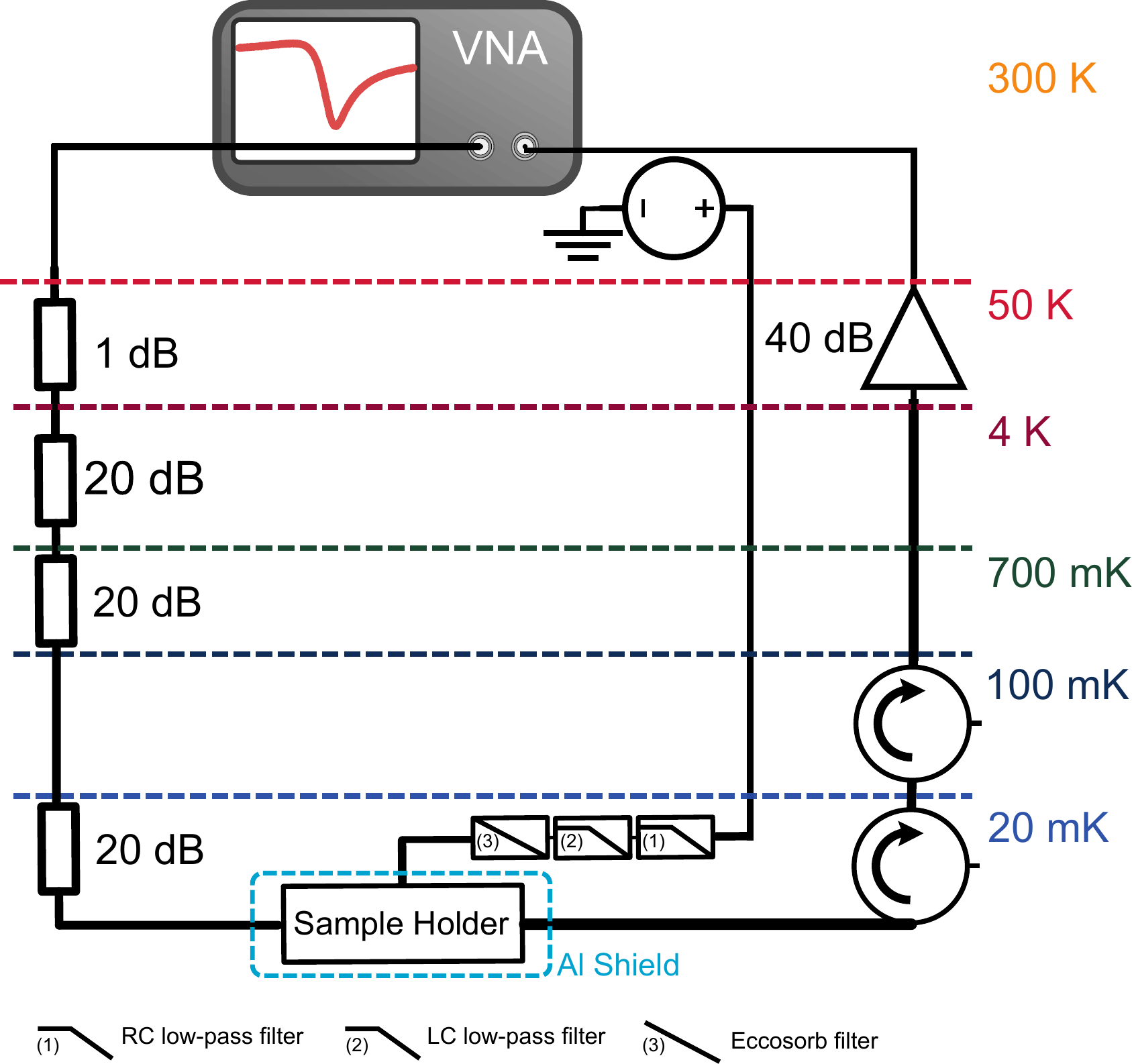}
    	    \caption{Schematic of the experimental setup used for one and two-tone spectroscopy measurements. RF measurements are performed using a vector network analyzer (VNA) to measure $S_{21}$ through the sample, though a series of cryogenic attenuators, and amplified through cryogenic and room-temperature amplifiers. The offset charge $N_g$ is controlled by a voltage source, connected through resistive thermocoax wiring to the mixing chamber. The voltage signal is further filtered through an RC (cutoff $20~\mathrm{kHz}$) and $LC$ (cutoff $80~\mathrm{MHz}$) and infra-red (eccosorb) filter.}
    	    \label{fig:experimentalsetup}
        \end{figure}

Measurements are performed in a cryogen-free dilution refrigerator with a base temperature of $20~\mathrm{mK}$. Using a VNA, a probe microwave tone is supplied to the feedline through a $80~\mathrm{dB}$ of attenuation distributed at the various temperature stages of the cryostat. The probe signal is then passed through two cryogenic circulators, before being amplified first by a $40~\mathrm{dB}$ cryogenic amplifier, and secondly by a $40~\mathrm{dB}$ room-temperature amplifier. The offset charge, $N_g$, is supplied by a nearby voltage gate, with DC component passed though an low temperature $RC$ filter, $LC$ filter, and eccoscorb filter, and connected to an isolated voltage source at room temperature. The device is mounted in a tight cooper holder and covered by an Al-shield to protect from stray magnetic field and incident radiation.

\section {$LCL$-Filter Transmission}\label{app:LCLFilter}
The $2\times2$ transmission matrix, or ABCD matrix, of a network is constructed by multiplying the ABCD matrices of each individual two-port element sequentially \cite{Pozar:882338}. In the case of $LCL$ network, the ABCD matrix is given by the multiplication of ABCD matrices of the inductor ($Z_{L}$) in series, capacitor ($Z_{C}$) in parallel and inductor ($Z_{L}$) in series, as schematically shown in Fig.~\ref{fig:3}(a) and formulated as follows
\begin{equation}\label{eq:ABCDMatrix}
\begin{pmatrix}
A & B
\\C & D
\end{pmatrix}
=
\begin{pmatrix}
1 & Z_L(f)
\\0 & 1
\end{pmatrix}
\begin{pmatrix}
1 & 0
\\1/Z_C(f) & 1
\end{pmatrix}
\begin{pmatrix}
1 & Z_L(f)
\\0 & 1
\end{pmatrix}
\end{equation}
The voltage ratio between port~2 and port~1 ($S_{21}$), can then be calculated as
\begin{equation}
S_{21}(f)=\frac{2}{A+B/Z_{0}+CZ_{0}+D}
\end{equation}

\begin{equation}\label{eq:S21Analytical}
S_{21}(f)=\frac{2Z_{0}Z_{C}}{2Z_{0}(Z_{L}+Z_{C})+(2Z_{L}Z_{C}+Z_{L}^{2})+Z_{0}^{2}}
\end{equation}
where $Z_{L}= j 2\pi f L_{LCL} $, $Z_{C}=j/(2\pi f C_{LCL}) $, and $Z_{0}=50~\Omega$; $L_{LCL}$ and $C_{LCL}$ are inductance and capacitance values of the $LCL$-filter. 

\section*{Acknowledgments}
We acknowledge Dr. Joonas Peltonen, Dr. Dmitry Golubev, Dr. George Thomas, Dr. Neill Lambert and Ilari Mäkinen for technical support and insightful discussions. This work is financially supported through the Foundational Questions Institute Fund (FQXi) via Grant No. FQXi-IAF19-06, Academy of Finland grants 312057, the Russian Science Foundation (Grant No. 20-62-46026) and from the European Union’s Horizon 2020 research and innovation programme under the European Research Council (ERC) (Grant No. 742559). We acknowledge the provision of facilities by Micronova Nanofabrication Centre and OtaNano - Low Temperature Laboratory of Aalto University to perform this research. We thank VTT Technical Research Center for sputtered Nb films.

\section*{Data Availability Statement}
All data used in this paper are available upon request to the authors, including descriptions of the data sets, and scripts to generate the figures.

\pagebreak

%

\end{document}